%% file: main.tex
\documentclass[prd, aps, showkeys, twocolumn, superscriptaddress, showpacs, nofootinbib, usenatbib, longbibliography]{revtex4-2}
\usepackage{graphicx} % Required for inserting images
\usepackage{natbib} 
\usepackage{enumitem}
\usepackage{hyperref}
\usepackage[table]{xcolor}
\usepackage{xcolor}
\usepackage{braket}
\usepackage[caption=false]{subfig}
\usepackage{pdfcomment}
\usepackage{todonotes}
\usepackage{nicematrix}
\usepackage{siunitx}[=v2]
\usepackage[nolist,nohyperlinks]{acronym}
\usepackage{amsmath, amsthm, amssymb, amsbsy, mathtools, mathalpha}
\usepackage{booktabs}

\renewcommand{\d}{{\bf d}}

\newcommand{\bmu}{\mbox{\boldmath $\mu$}}
\newcommand{\bxi}{\mbox{\boldmath $\xi$}}
\newcommand{\bphi}{ \phi}
\newcommand{\btheta}{\mbox{\boldmath $\theta$}}
\newcommand{\bdelta}{\mbox{\boldmath $\delta$}}

\newcommand{\Z}{\mathbf{Z}}
\renewcommand{\d}{\mathbf{d}}
\renewcommand{\S}{\mathbf{S}}
\newcommand{\tr}{\mbox{tr}}
\newcommand{\I}{\mbox{{\bf I}}}
\newcommand{\e}{{\rm e}}

\renewcommand{\u}{\mathbf {u}}
\renewcommand{\v}{{\bf v}}
\newcommand{\Y}{\mbox{{\bf Y}}}

\usepackage{dcolumn}% Align table columns on decimal point
\usepackage{bm}% bold math
\usepackage{color, units}
\usepackage[normalem]{ulem}

\usepackage{xspace}
\usepackage{comment}

\begin{document}
\title{Variational Bayesian Inference for \\ the Spectral Structure of LISA Noise}

\input{authors}
\date{\today}

\begin{abstract}
Estimating spectral density matrices for future space-based gravitational-wave detectors such as LISA is challenging due to the long duration of the data and the correlated instrumental noise across multiple time-delay interferometry channels. In this work, we investigate a specialized mean-field stochastic gradient variational Bayes (SGVB) procedure for fast posterior approximation in long duration multivariate spectral density estimation. Based on the existing eigenbasis representation of
the blocked Whittle likelihood approach, the posterior model applies a Cholesky factorization to represent the inverse spectral density matrix and models the resulting frequency-dependent entries with cosine basis functions, with a discounted regularized horseshoe prior assigned to the basis coefficients.
We test mean-field SGVB as a stand-alone posterior approximation by comparing it with Hamiltonian Monte Carlo (HMC) targeting the same posterior model. In simulation studies based on autoregressive and moving average processes, SGVB produces posterior median spectral estimates close to those obtained from HMC at substantially lower computational cost. We then apply the method to two one-year, fixed-delay, stationary, noise-only LISA  simulations and show that the SGVB spectral density estimates are consistent with HMC and Welch estimates across the LISA analysis band, while requiring substantially less computation. These results demonstrate that SGVB provides a scalable Bayesian approach to spectral density estimation for long-duration multivariate LISA noise analysis.
% \avi{[Discuss maybe] The abstract currently reads as if the pipeline is new, but the components already exist across Hu et al.\ 2023, our ET paper, Liu et al.\ 2025, and our submitted multivariate P-spline paper. In particular, the P-spline paper already derives and uses this same eigendecomposition of the summed periodogram. Do not present the eigenbasis itself as the novelty. The abstract must instead state the genuinely distinct contribution of this paper. at present i think that seems to be the specialized mean-field SGVB implementation as a fast stand-alone posterior approximation for year-long LISA data, benchmarked against HMC... is that sufficient novelty for a separate PRD paper? I think our novelty needs to be made explicit and defended? } \Jianan{The abstract is modified based on Avi's suggestions}
\end{abstract}

\maketitle

% Acronyms
\begin{acronym}
    \acro{GW}[GW]{gravitational-wave}
    \acro{LISA}[LISA]{Laser Interferometer Space Antenna}
    \acro{PSD}[PSD]{power spectral density}
    \acro{DFT}[DFT]{Discrete Fourier Transform}
    \acro{SVI}[SVI]{stochastic variational inference}
    \acro{KL}[KL]{Kullback-Leibler}
    \acro{MCMC}[MCMC]{Markov chain Monte Carlo}
    \acro{TDI}[TDI]{time-delay interferometry}
    \acro{LVK}[LVK]{LIGO--VIRGO--KAGRA}
    \acro{SGVB}[SGVB]{Stochastic gradient variational Bayes}
    \acro{HMC}[HMC]{Hamiltonian Monte Carlo}
    \acro{NUTS}[NUTS]{No-U-Turn Sampler}
    \acro{OMS}[OMS]{optical metrology system}
    \acro{TM}[TM]{test-mass}
    \acro{MOSA}[MOSA]{movable optical sub-assembly}
    \acro{VAR}[VAR]{vector autoregressive}
    \acro{VMA}[VMA]{vector moving-average}
    \acro{RISE}[RISE]{relative integrated squared error}
    \acro{RIAE}[RIAE]{relative integrated absolute error}
    \acro{MAE}[MAE]{mean absolute error}
    \acro{ESS}[ESS]{effective sample size}
\end{acronym}

\section{Introduction}

Reliable estimation of the instrumental noise \ac{PSD} is a central challenge for space-based  \ac{GW} detectors such as the \ac{LISA}. In many analyses of the current \ac{LVK}
detector network, the noise in spatially separated detectors is often treated as independent. By contrast, LISA observations are commonly represented by three \ac{TDI} Michelson channels~\citep{Amaro2017}. The instrumental noise in these channels can exhibit frequency-dependent cross-spectral structure because the TDI Michelson channels are built from shared inter-spacecraft phase measurements, so common link-level noise contributions can enter multiple channels through different transfer functions. Separate single-channel PSDs therefore cannot provide a complete description of the joint noise behavior across channels.

Neglecting the cross spectrum is equivalent to treating the TDI
channels as independent at each frequency. \citet{Boileau2021} adopted a parametric Bayesian model for the instrumental-noise PSD and used Adaptive \ac{MCMC} to infer the associated noise parameters jointly with stochastic-background parameters, under a diagonal spectral covariance assumption for the TDI channels. Such a simplification can misrepresent the joint noise behavior between TDI channels and may affect downstream analyses, including stochastic-background searches and gravitational-wave parameter estimation~\citep{Cireddu2024, Hartwig2023}. Therefore, LISA noise estimation requires methods that can estimate a Hermitian positive-definite spectral density matrix at each frequency, thereby capturing the joint spectral behavior of the TDI channels.

A classical nonparametric approach is Welch estimation, which averages periodograms over data segments to obtain the noise spectrum~\citep{welch1967}, but it returns a point estimate rather than posterior uncertainty quantification. A common class of LISA noise-inference approaches specifies the spectral covariance of the TDI observables by propagating link-level noise spectra through prescribed TDI transfer functions \citep{Nam2023,Baghi2023,Santini2025,liang2026fullcovariancebayesianinferencestochastic}. Although the link-level spectra can be modeled flexibly, for example using spline representations \citep{Baghi2023,Santini2025}, these approaches typically treat the transfer functions as fixed and impose structural assumptions on the link noises, such as a common spectrum across individual links \citep{Baghi2023}. The resulting TDI PSDs and cross-spectra are therefore constrained by the assumed response and link-noise structure.

By directly estimating the full multivariate spectral density matrix at the TDI level, Bayesian nonparametric methods provide a more flexible alternative. This formulation can accommodate the aggregate contribution of all potential noise sources without treating the TDI transfer functions as deterministic and known. Many existing approaches perform posterior inference through MCMC sampling~\cite{RosenOri2007Aeom, Meier2020, Liu2024, Guillermo2024}. Although these methods provide flexible inference for multivariate spectra, their computational cost can become substantial for long time series. To improve scalability, variational inference approximates the posterior by optimizing the parameters of a tractable variational distribution, thereby reformulating posterior inference as an optimization problem~\cite{Blei2017,Kingma2013,Hoffman2012}. \ac{SGVB} further enables this optimization to be carried out using stochastic gradient estimators and reparameterized Gaussian approximations \citep{Kingma2013,Xu2019,Domke2019}.

Several recent methods are particularly closely related to the present multivariate analysis. Building on the SGVB estimator of \citet{Hu2023} and the blocked Whittle likelihood of \citet{Liu2025}, \citet{Vajpeyi2026} parameterize the Cholesky components of the inverse spectral matrix with penalized B-splines and hierarchical smoothing priors. They combine a blocked, coarse-grained Whittle likelihood with safe-Bayes tempering and use the \ac{NUTS} initialized by \ac{SVI} for posterior inference. \citet{Liu2026} instead represent the spectral matrix using a Bernstein polynomial expansion with matrix-valued coefficients assigned a matrix-gamma process prior, and perform posterior inference using adaptive MCMC. They also develop a semiparametric extension in which the nonparametric prior models a correction to a parametric working spectrum. Both approaches directly estimate a Hermitian positive-definite spectral density matrix and demonstrate their methods on simulated gravitational-wave detector noise.

In this work, we present a more computationally efficient Bayesian method to estimate multivariate PSD matrices from long stationary time series. Similar to the P-spline approach \citep{Vajpeyi2026}, we adopt the same eigenbasis representation of the blocked multivariate Whittle likelihood, but using cosine basis expansions~\citep{RosenOri2007Aeom,Hu2023} to represent the components obtained from the Cholesky factorization of the inverse spectral density matrix. A discounted regularized horseshoe prior is assigned to the corresponding basis coefficients~\citep{PiironenJuho2017Siar,Hu2023}. SGVB\citep{Hu2023} is used as the primary posterior approximation, rather than only as a preliminary step before sampling. We benchmark SGVB against \ac{HMC} under the same posterior model, focusing on whether SGVB can provide accurate spectral estimates at substantially lower computational cost.

Through simulation studies based on \ac{VAR} models of order 2 and \ac{VMA} models of order 1 with known true PSD matrices, we illustrate the performance of SGVB and compare it to HMC.%, we show that the posterior median estimates from SGVB are close to those obtained from HMC, while SGVB requires substantially less computation time. However, the empirical coverage probabilities of the credible intervals for SGVB are generally lower than those of HMC. 
We further apply the method to two one-year simulated LISA TDI datasets, \texttt{noise4a} \citep{lisa_noise4a} and \texttt{noise5a} \citep{lisa_noise5a}. Both are fixed-delay, stationary, noise-only cases and therefore exclude gravitational-wave signals, orbital breathing, and time-dependent arm-length modulation. The \texttt{noise4a} dataset has symmetric \ac{OMS} and \ac{TM} noise levels across the relevant optical paths, whereas \texttt{noise5a} introduces asymmetry through independently scaled \acs{OMS} and \acs{TM} noise amplitudes. 
%The resulting SGVB posterior median estimates are consistent with the HMC and Welch estimates across the LISA analysis band, as measured by the \ac{RIAE}.

The paper is structured as follows. Section~\ref{sec:method} introduces an eigendecomposition-based representation of the blocked multivariate Whittle likelihood, the proposed Bayesian spectral density model, and the corresponding SGVB inference procedure. Section \ref{sec:simulation} evaluates the proposed method through simulation studies, assessing the accuracy and computational efficiency of SGVB. Section~\ref{sec:applications} applies the proposed method to one-year simulated LISA TDI noise data and compares the resulting multivariate PSD estimates with those obtained using HMC. Section~\ref{sec:discussion} concludes the paper with a discussion of the main findings and future research directions.
% \avi{[Easy] Housekeeping for the whole draft (listing once here): (v) Mixed \texttt{\textbackslash citep}/\texttt{\textbackslash cite} usage -- pick one convention for revtex. (vi) Check \texttt{Karnesis2026} -- a 2026 reference should be verified as published/posted.}
% \Jianan{I used \texttt{cite} in places where several references are grouped together, because it compresses consecutive numerical references into a range. I have removed the citation to Karnesis2026 for now.}

\section{Method}
\label{sec:method}
\subsection{Likelihood}
We denote a  $p$-dimensional stationary, mean-zero multivariate time series as $\Z=(\Z_1,\ldots,\Z_n)^ \intercal\in  \mathbb{R}^{n\times p}$, sampled at time intervals $\Delta_t=1/(2f_{Ny})$, so that $\Z_t=\Z(t\Delta_t)\in \mathbb{R}^p$ for $t=1,\ldots,n$, where $f_{Ny}$ is the Nyquist frequency. The total observation time is $T$, each dimension contains $n=T/\Delta_t$ sampled values. The corresponding frequency resolution is
\begin{align}
\Delta_f = \frac{1}{n \Delta_t} = \frac{1}{T}\, .
\end{align}
The discrete Fourier transform (DFT) of $\Z$ is
\begin{align}
\d(\nu_k) = \Delta_t\sum_{t=1}^{n} \Z_t\exp \left(-2\pi i \frac{k}{n} t \right)\, ,
\end{align}
where the Fourier frequencies are
$\nu_k= k \Delta_f= k/({n\Delta_t})=k /{T}$ for $k=1,\ldots, N$. Here,
$N=n/2$ when $n$ is even, and $N=(n-1)/2$ when $n$ is odd.
For stationary time series with absolutely summable autocovariances $\sum_{h=-\infty}^{\infty}||\Gamma(h)||<\infty$, the discrete Fourier coefficients $\d(\nu_k)$ are asymptotically independent and distributed as complex Gaussian random vectors with mean zero and covariance matrix $T\,\S(\nu_k)$, where $\S(\nu_k)$ denotes the two-sided spectral density matrix evaluated at the $k$th Fourier frequency,
\begin{equation}
\S(\nu_k)=\frac{1}{2f_{Ny}} \sum_{h=-\infty}^{\infty} \Gamma(h) \exp \left(-2\pi i \nu_k h\Delta_t \right),
\end{equation}
i.e., the Fourier transform of the time-invariant autocovariance $\Gamma(h)=\mathbb{E}(\Z_{t}\Z_{t+h}^\intercal)$. This asymptotic complex Gaussian approximation leads to the multivariate Whittle likelihood in the frequency-domain, which can be expressed as
\begin{align}\label{eq:Whittle likelihood}
 \mathcal{L}(\d|\S) &\propto  \prod_{k=1}^{N} \det(\S(\nu_k))^{-1} \times \nonumber \\
 & \exp\left(-\frac{1}{T}\d(\nu_k)^* \S(\nu_k)^{-1} \d(\nu_k)\right),
\end{align}
where $\d(\nu_k)^*$ represents the conjugate transpose of $\d(\nu_k)$ and $\S(\nu_k)$ is a $p \times p$ Hermitian positive definite spectral density matrix at each $\nu_k$.

When the time series length $n$ is large, direct evaluation of the multivariate Whittle likelihood over all Fourier frequencies can be computationally expensive. To make the inference computationally feasible, a blocked Whittle likelihood is adopted \citep{Liu2025}. The observed series is divided into $N_b$ non-overlapping equal-sized blocks $\Z = \left(\Z^{(1)},\ldots,\Z^{(N_b)}\right)^\intercal$, where each $\Z^{(i)}$ is a $p$-dimensional time series of length $n_b=n/N_b$, and each block has duration $T_b = T/N_b$. Non-overlapping blocks are used so that the block DFTs can be treated as approximately independent under the blocked Whittle likelihood. For each block, the Fourier frequencies are defined as $f_k={k}/{T_b}$, where $ k=1, \ldots,n_b/2$.
Let $\d^{(i)}(f_k)$ denote the DFT of the $i$th block $\Z^{(i)}$ at frequency $f_k$, for $i=1,\ldots,N_b$. Under stationarity, all blocks share the same spectral density matrix. With the additional approximation that different blocks are independent, the blocked Whittle likelihood is obtained by multiplying the Whittle likelihood contributions from the individual blocks,
\begin{equation}\label{eq:block_lnl}
    \mathcal{L}_b(\d|\S) = \prod^{N_b}_{i=1} \mathcal{L}(\d^{(i)}|\S) \ .
\end{equation}
Define the periodogram matrix for the $i$th block at frequency $f_k$ as $\I^{(i)}(f_k)=\d^{(i)}(f_k)\d^{(i)}(f_k)^{*}$, then summing these matrices over all blocks gives 
\begin{equation}
\Y(f_k)=\sum_{i=1}^{N_b}\I^{(i)}(f_k).
\end{equation}
Following \citet{Liu2026}, and using the cyclic invariance of the trace, for matrices with compatible dimensions, the trace satisfies $\tr(ABC) = \tr(CAB)$, then the blocked Whittle likelihood can be written explicitly as
\begin{align}
\mathcal{L}_b(\d|\S) &\propto  \prod_{k=1}^{n_b/2}  \left|\S(f_k)\right|^{-N_b} \times \notag \\
&\exp\left(-\frac{1}{T_b} \tr\left[ \S(f_k)^{-1} \Y(f_k)\right]\right).
\end{align}
At each frequency $f_k$, $\Y(f_k)$ is a $p \times p$ Hermitian positive semidefinite matrix, and it has a $p-$dimensional complex Wishart distribution:
\begin{align}
\Y(f_k)\sim \mathcal{CW}_p(T_b\S(f_k), N_b).
\end{align}

Since $\Y(f_k)$ is a Hermitian positive semidefinite matrix, it admits an eigendecomposition with non-negative eigenvalues. We adopt the eigenbasis representation of the blocked Whittle likelihood used in \citet{Vajpeyi2026}. Let $\lambda_\nu^{(k)}$ and $\v_ \nu^{(k)}, \nu=1,\ldots,p$, denote the eigenvalues and the corresponding eigenvectors of $\Y(f_k)$, then
\begin{equation}
\Y(f_k) = \sum_{\nu=1}^{p}\lambda^{(k)}_{\nu}\v^{(k)}_{\nu}\v^{(k)*}_{\nu} = \sum_{\nu=1}^{p}\u^{(k)}_{\nu}\u^{(k)*}_{\nu},    
\end{equation}
where the eigenvectors are rescaled as $\u^{(k)}_{\nu}=\sqrt{\lambda^{(k)}_{\nu}}\v^{(k)}_{\nu} \in  \mathbb{C}^{p}$.
Substituting this decomposition into the above expression for the blocked Whittle likelihood $\mathcal{L}_b(\d|\S)$ gives
\begin{align} \label{eigenbasis_likelihood}
    \mathcal{L}_b(\d|\S)
    &\propto
    \prod_{k=1}^{n_b/2} \left|\S(f_k)\right|^{-N_b} \times \notag \\
    &\exp\left(
    -\frac{1}{T_b}
    \sum_{\nu=1}^{p} \u_{\nu}^{(k)*}\S(f_{k})^{-1}\u_{\nu}^{(k)}
    \right).
\end{align}
%In this form, the trace term involving $\Y(f_k)$ is rewritten as a sum of vector quadratic forms involving the rescaled eigenvectors $\u_\nu^{(k)}$, this avoids evaluating the trace with the full aggregated periodogram matrix directly. These quadratic forms can also be evaluated separately over $\nu$, making the likelihood well suited for parallel computation. Hence, the likelihood is more computationally efficient to evaluate.
% \avi{I think this computational argument is slightly mis-stated. For $p=3$, evaluating $\tr[\S^{-1}\Y]$ is trivial. the useful role of the eigendecomposition is to express the Wishart sufficient statistic as vector quadratic forms compatible with the component-wise Cholesky regressions, right? Also -- it is exact, not a new likelihood approximation. However, this same derivation and implementation already appear in our submitted multivariate P-spline paper, so describe it as an adopted computational representation and cite that paper rather than presenting it as this paper's central methodological contribution, because that would be misleading. The paper's speed claim should be framed as SGVB versus HMC for this posterior, not as evidence that the eigendecomposition itself is new.}
% \Jianan{I have added the citation at the beginning of this paragraph. The factorized likelihood with Cholesky component will be shown in next section}

In practice, in order to reduce spectral leakage caused by the finite block length, each block $\Z^{(i)}$ may be tapered by a window function $\omega_t$ before applying the DFT. However, tapering induces correlation between neighboring Fourier bins and therefore reduces the effective number of independent frequency-domain observations. To account for this loss of effective information, the blocked Whittle log-likelihood is conservatively divided by the normalized equivalent noise bandwidth $N_{\rm bw}$, where
\begin{equation}
N_{\rm bw} =
\frac{
n_b\sum_{t=1}^{n_b}\omega_t^2
}{
\left(\sum_{t=1}^{n_b}\omega_t\right)^2
},
\label{eq:normalized-enbw}
\end{equation}
$N_{\rm bw}=1$, for a rectangular window. Equivalently, the taper-adjusted blocked Whittle likelihood can be written as
\begin{equation}\label{nbw}
\mathcal L_{b,\omega}(\d|\S) \propto
{\mathcal L_{b}(\d|\S)}^{1/N_{\rm bw}}.
\end{equation}
The fractional power in Equation~\ref{nbw} can be considered a form of likelihood tempering, following the same general idea as safe Bayes adjustments for approximate or misspecified likelihoods \citep{Grunwald2014,Vajpeyi2026}. In the present work, the tempering exponent is determined by the normalized equivalent noise bandwidth. The choice of taper and the resulting value of $N_{\rm bw}$ are specified for each analysis below.

\subsection{Parametrization of $\S$}
\label{subsec: parameters}
To ensure the estimated spectral density matrix $\S(f_k)$ is positive definite at each frequency, we adopt the parametrization of \citet{RosenOri2007Aeom} and \citet{Hu2023}, in which the inverse spectral density matrix is represented through Cholesky decomposition and its components are modelled by basis functions. 
The inverse of $\S(f_k)$ can be decomposed as $\S(f_k)^{-1} = \mathbf{T}_k^* \mathbf{D}_k^{-1} \mathbf{T}_k$, where $\mathbf{D}_k$ is a diagonal matrix with diagonal elements $\delta_{1k}^2, \delta_{2k}^2, ..., \delta_{pk}^2$ and
\begin{align}
\mathbf{T}_k = \begin{pmatrix}
1 & 0 & 0 & \cdots & 0 \\
-\theta_{21}^{(k)} & 1 & 0 & \cdots & 0 \\
-\theta_{31}^{(k)} & -\theta_{32}^{(k)} & 1 & \ddots & \vdots \\
\vdots & \vdots & \ddots & \ddots & 0 \\
-\theta_{p1}^{(k)} & -\theta_{p2}^{(k)} & \cdots & -\theta_{p,p-1}^{(k)} & 1
\end{pmatrix}
\end{align}
is a $p \times p$ complex unit lower triangular matrix. Using the Cholesky parametrization, the blocked eigenbasis likelihood can be factorized into $p$ componentwise contributions,
\begin{equation}
\mathcal{L}_{b}(\d|\S) \propto \prod_{j=1}^{p}\mathcal{L}_{b,j}(\u_j|{\btheta_j,\bdelta_j},\u_{<j}),
\end{equation}
where
\begin{align}
\label{component_likelihood}
\mathcal{L}_{b,j}(\u_j|{\btheta_j,\bdelta_j},\u_{<j}) &\propto
\prod_{k=1}^{n_b/2} \delta_{jk}^{-2N_b} \times \notag \\
&\exp\left(
\frac{-\sum_{\nu=1}^{p}\left|u_{j\nu}^{(k)}-\sum_{l=1}^{j-1}\theta_{jl}^{(k)}u_{l\nu}^{(k)} \right|^2}{T_b\delta_{jk}^2}
\right),
\end{align}
$u_{j\nu}^{(k)}$ denotes the $j$th component of the $\nu$th rescaled eigenvector at each frequency bin, $\u_{j}$ collects these entries over all frequencies and eigenvector indices, while $\u_{<j}$ denotes the corresponding entries with component indices smaller than $j$. The parameters $\btheta_j,\bdelta_j$ represent the set of $\theta_k$ and $\delta_k$ for the $j$th component of the multivariate time series, $\theta_{jl}^{(k)}$ represents the corresponding element in the matrix $\mathbf{T}_k$ for $j>l$. The Cholesky parametrization introduces an ordering of the components, so the componentwise likelihood is defined relative to this fixed ordering.
For $p=3$ (as in the LISA case), Equation~\ref{component_likelihood} gives the following
three componentwise likelihood contributions for $j=1,2,3$,
\begin{align}
\mathcal{L}_{b,1}(\u_1|{\bdelta_1}) \propto
\prod_{k=1}^{n_b/2} \delta_{1k}^{-2N_b} \exp\left(
\frac{-\sum_{\nu=1}^{p}\left|u_{1\nu}^{(k)} \right|^2}{T_b\delta_{1k}^2}
\right),
\end{align}

\begin{align}
\mathcal{L}_{b,2}(\u_2|{\btheta_2,\bdelta_2},\u_{1}) &\propto 
\prod_{k=1}^{n_b/2} \delta_{2k}^{-2N_b} \times \notag \\
& \exp\left(
\frac{-\sum_{\nu=1}^{p}\left|u_{2\nu}^{(k)} -\theta_{21}^{(k)}u_{1\nu}^{(k)} \right|^2}{T_b\delta_{2k}^2}
\right),
\end{align}

\begin{align}
&\mathcal{L}_{b,3}(\u_3|{\btheta_3,\bdelta_3},\u_{1},\u_{2}) \notag \\ 
&\propto
\prod_{k=1}^{n_b/2} \delta_{3k}^{-2N_b}
\exp\left(
\frac{-\sum_{\nu=1}^{p}\left|u_{3\nu}^{(k)} -\theta_{31}^{(k)}u_{1\nu}^{(k)} - \theta_{32}^{(k)}u_{2\nu}^{(k)}\right|^2}{T_b\delta_{3k}^2}
\right).
\end{align}
% \avi{I think several fixes are needed. (d) Worth one sentence acknowledging if or if not the Cholesky factorization depends on the channel ordering (X,Y,Z here), a standard caveat of this parametrization that a careful referee will raise (i dont think it matters in our case -- i had checked for pspline paper).}
% \Jianan{The sentence is added for ordering matter.}
Then $\log(\delta_{jk}^2)$ and the real and imaginary parts of $\theta_{jl}^{(k)}$ are modeled by Demmler-Reinsch basis functions, given by
\begin{align}
\Re(\theta_{jl}^{(k)}) &= \alpha_{jl,0} + \alpha_{jl,1}f_k + \sum_{s=1}^{M-1}\psi_s(f_k)\alpha_{jl,s+1}, \label{eq:real_theta} \\
\Im(\theta_{jl}^{(k)}) &= \beta_{jl,0} + \beta_{jl,1}f_k + \sum_{s=1}^{M-1}\psi_s(f_k)\beta_{jl,s+1}, \label{eq:imag_theta} \\
\log \delta_{jk}^2 &= \gamma_{j,0} + \gamma_{j,1}f_k + \sum_{s=1}^{M-1}\psi_s(f_k)\gamma_{j,s+1},  \label{eq:log_delta}
\end{align}
where $\psi_s(f_k) = \sqrt{2} \cos(s\pi \frac{f_k}{2f_{Ny}})$ denotes the $s$-th basis function, for $s=1,...,M-1$.
% \avi{And also for LISA we fit over $10^{-4}$--$10^{-1}$~Hz spanning three decades -- are the basis functions evaluated on a linear or log frequency axis? With linear scaling, virtually no resolution lands in the lowest decade... which likely explains the low-frequency sensitivity to $M$ seen in the appendix. this needs to be stated and ideally discussed.}
% \Jianan{The cosine basis functions are evaluated on a normalized linear frequency scale. I have added a short discussion of this point in Appendix B.}
Thus, $M-1$ basis functions are included in addition to the intercept and linear terms. The flexibility of the spectral density estimate is controlled by the number of basis functions $M$. The basis functions $\psi_s$ are ordered from very smooth to increasingly wiggly, and the corresponding coefficients ($\alpha,\beta,\gamma$) determine the weights of each basis function. We adopt a flexible prior structure for the spectral densities by placing shrinkage priors on the basis coefficients.  Following \citet{Hu2023}, we assign a discounted regularized horseshoe prior \citep{PiironenJuho2017Siar} to these basis coefficients. This prior combines global and local shrinkage with an additional discounting factor that imposes stronger shrinkage on higher order basis coefficients. Therefore, it provides a data-adaptive balance between smoothness and flexibility. The hierarchical form of the prior is given in Appendix \ref{priors}. 
Let $\boldsymbol{\eta}$ denote the full parameter vector comprising the coefficient vectors $\boldsymbol{\alpha}$, $\boldsymbol{\beta}$, and
$\boldsymbol{\gamma}$, together with the corresponding prior hyperparameters.
% \avi{For a PRD audience, give the hierarchical form of the discounted regularized horseshoe (even if just in an appendix) rather than pointing entirely to Hu et al. This paper should be self-contained on its own prior, and ``discounted'' is never defined. Also, i think there is a discrepancy to address: this prior is invoked to ``avoid overfitting'', yet Appendix~A shows the fits degrade visibly for large $M$, which is exactly the overfitting the horseshoe is supposed to suppress?? Im confused... Reconcile these (see my new comment in the appendix).}
% \Jianan{I have written a section for priors in appendix. I have aovided the characters as avoiding overfitting.}
Partition $\boldsymbol{\eta}$ into $p$ subvectors as $\boldsymbol{\eta} = (\boldsymbol{\eta}_1,\ldots,\boldsymbol{\eta}_p)$, where each $\boldsymbol{\eta}_j$ contains the basis coefficients and prior hyperparameters associated with the $j$th likelihood contribution. Independent priors are assigned to these subvectors, the joint prior for $\boldsymbol{\eta}$ can be factorized as
\begin{align}
\pi(\boldsymbol{\eta}) =
\prod_{j=1}^{p}\pi_j(\boldsymbol{\eta}_j).
\end{align}
Therefore, the posterior distribution is given by Bayes' theorem as
\begin{align}
    p(\boldsymbol{\eta}|\d) =& \frac{\mathcal{L}_{b,w}(\d|\boldsymbol{\eta})\pi(\boldsymbol{\eta})}{\mathcal{Z}(\d)} \nonumber \\
    \propto& \mathcal{L}_{b,w}(\d|\boldsymbol{\eta})\pi(\boldsymbol{\eta})\
\end{align}
where $p(\boldsymbol{\eta} \mid \d)$ denotes the posterior density and $\mathcal{Z}(\d)$ is the Bayesian evidence.
For the three-channel setting considered in the LISA application, the posterior density can be expressed as
\begin{align}
p(\boldsymbol{\eta}|\d) \propto {}&
\mathcal{L}_{b,w,1}(\u_1|\bdelta_1) \pi_1(\boldsymbol{\eta}_1) \nonumber\\ &\times
\mathcal{L}_{b,w,2}(\u_2|\btheta_2,\bdelta_2,\u_1) \pi_2(\boldsymbol{\eta}_2) \nonumber\\ &\times
\mathcal{L}_{b,w,3}(\u_3|\btheta_3,\bdelta_3,\u_1,\u_2) \pi_3(\boldsymbol{\eta}_3).
\label{eq:component_posterior}
\end{align}
Here, $\mathcal{L}_{b,w,j} = \mathcal{L}_{b,j}^{1/N_{\mathrm{bw}}}$ denotes the taper-adjusted  likelihood contribution for component $j$.

\subsection{Stochastic Gradient Variational Bayes}
\label{subsec:sgvb_details}
The main idea of variational inference is to approximate the posterior density $p(\boldsymbol{\eta}|\d)$ by optimizing the parameters of a surrogate distribution from a tractable family of variational distributions ${\cal Q}=\{ q_{\phi}(\boldsymbol{\eta}) : \phi \in \Phi \}$, where $\phi$ denotes the variational parameter vector for a particular member of the family, and $\Phi$ is the corresponding variational parameter space. The surrogate distribution is optimized when the \ac{KL} divergence between the resulting distribution and the true posterior distribution is minimized, i.e.,
\begin{align}\label{eq:phi_min}
  \bphi_j^* &= \operatorname*{argmin_{\bphi_j}} d_{KL}(q_{\bphi_j}||p_j) \nonumber \\
  &= \operatorname*{argmin_{\bphi_j}} \int \log\frac{q_{\bphi_j}(\boldsymbol{\eta}_j)}{p_j(\boldsymbol{\eta}_j|\d_j,\d_{<j})}q_{\bphi_j}(\boldsymbol{\eta}_j) \text{d}\boldsymbol{\eta}_j \, .
\end{align}
where $q_{\bphi_j}(\boldsymbol{\eta}_j)$ denotes the
variational approximation to the $j$th component of the posterior, while $p_j(\boldsymbol{\eta}_j|\d_j,\d_{<j})$ denotes the corresponding posterior component, $d_{KL}(q_{\bphi_j}||p_j)$ denotes the KL divergence. In this way, variational Bayes replaces the challenging problem of direct posterior integration with an optimization problem.

In this study, a stochastic gradient variational Bayes approach is utilized to approximate the posterior distribution of the spectral density parameters.
A multivariate Gaussian distribution with a diagonal covariance matrix is adopted as the surrogate distribution. The corresponding variational parameters for each $j$th component are denoted by ${\bphi}_j=\{\boldsymbol{\mu}_j,\boldsymbol{\xi}_j\}$, where $\boldsymbol{\mu}_j$ is the variational mean vector and $\boldsymbol{\xi}_j$ contains the logarithms of the diagonal entries in the covariance matrix. 
The optimization procedure for the variational parameters $\bphi_j$ is implemented following the first two phases of the three-phase Variational Bayes (TPVB) algorithm in \citep{Hu2023}:
\begin{enumerate}
\item The log joint density, $\log p(\boldsymbol{\eta}_j,\d_j,\d_{<j})$ is maximized with respect to $\boldsymbol{\eta}_j$ using gradient ascent, yielding the maximum a posteriori (MAP) estimate, i.e.,
$\hat{\boldsymbol{\eta}}_j=\operatorname*
{argmax_{\boldsymbol{\eta}_j}} \log p(\boldsymbol{\eta}_j,\d_j,\d_{<j}),$ where $p(\boldsymbol{\eta}_j,\d_j,\d_{<j})$ denotes the unnormalized posterior. Then the variational mean is initialized as $\hat{\bmu}_j=\hat{\boldsymbol{\eta}}_j$. The MAP estimate is obtained using the \texttt{Adam} optimizer \citep{Adam2014}.

\item  Given $\hat{\bmu}_j$ from the first step, the evidence lower bound (ELBO) between $q_{\bphi_j}(\boldsymbol{\eta}_j)$ and $p(\boldsymbol{\eta}_j,\d_j,\d_{<j})$ i.e.,
\begin{align}\label{eq:elbo}
 \text{EL}&\text{BO}(q_{\bphi_j}(\boldsymbol{\eta}_j),\ p_j(\boldsymbol{\eta}_j|\d_j,\d_{<j}))  \nonumber\\
 &=\mathbb{E}_{\boldsymbol{\eta}_j\sim q_{\bphi_j}(\boldsymbol{\eta}_j)}[\log p_j(\boldsymbol{\eta}_j,\d_j, \d_{<j})-\log q_{\bphi_j}(\boldsymbol{\eta}_j)]
\end{align}
is maximized with respect to $\boldsymbol{\xi}_j$ via gradient ascent i.e., $\hat{\bxi}_j = \operatorname*
{argmax_{\xi_j}} \mathbb{E}_{\boldsymbol{\eta}_j\sim q_{\hat{\mu}_j, {\xi}_j}}[\log p(\boldsymbol{\eta}_j,\d_j,\d_{<j})-\log q_{\hat{\mu}_j,\xi_j}(\boldsymbol{\eta}_j)].$ The ELBO is maximized using the \texttt{Adam} optimizer. At each iteration, the stochastic ELBO gradient is estimated using one draw from the current surrogate distribution.
\end{enumerate}
The TPVB algorithm in \citep{Hu2023} also includes an optional fine-tuning phase 3, in which both $\hat{\bmu}_j$ and $\hat{\bxi}_j$ obtained from phase 2 are further updated by maximizing the ELBO. Equivalently, this step further minimizes the KL divergence between the posterior and surrogate distribution. The variational parameters optimized by phase 2 are close to their values after phase 3, and the additional updates did not materially improve the optimization criterion. Therefore, the present implementation omits the optional third phase for computational efficiency. The fine-tuning step can be added when further refinement of the variational approximation is required.
The posterior model and both optimization stages are implemented using \texttt{TensorFlow} \citep{tensorflow} and \texttt{Tensorflow-Probability} \citep{tensorflowProb}.

The two-phase procedure is performed sequentially to approximate the corresponding $p_j(\boldsymbol{\eta}_j, \d_j,\d_{<j})$ for $j=1,\ldots,p$. Both stages are run for fixed numbers of iterations without an additional convergence-based stopping rule.
After optimization, we independently generate 500 samples from each optimized surrogate distribution $q_{\hat{\boldsymbol{\mu}}_j,\hat{\boldsymbol{\xi}}_j} (\boldsymbol{\eta}_j)$ in both the simulation study and the LISA analyses, combining samples with the same index across components to form realizations of the full parameter vector $\boldsymbol{\eta}$. In each $\boldsymbol{\eta}$, the basis coefficients determine the components of the Cholesky representation, as defined in Equations~\eqref{eq:real_theta}, \eqref{eq:imag_theta}, and~\eqref{eq:log_delta}, which are then used to construct the corresponding PSD matrices across the frequency grid. At each frequency, posterior medians and credible intervals are obtained by summarizing the spectral density matrices based on all samples of $\boldsymbol{\eta}$.
This avoids sampling directly from the posterior distribution using any MCMC methods, leading to a substantial improvement in computational efficiency.

To assess the accuracy and computational efficiency of the SGVB approximation, a single-chain HMC sampler is also applied sequentially to $p_j(\boldsymbol{\eta}_j,\d_j,\d_{<j})$, for $j=1,\ldots,p$, using the \texttt{HamiltonianMonteCarlo} kernel in \texttt{TensorFlow-Probability}. For both the simulation study and LISA analyses, each chain is initialized at the corresponding phase 1 estimate $\hat{\boldsymbol{\eta}}$.
Each sampler uses an initial step size of $10^{-4}$ and performs 15 leapfrog steps at each HMC update. During the first $80\%$ of the burn-in period, the step size is adapted to target an acceptance probability of $0.85$. All SGVB and HMC analyses were performed using a single CPU core on the New Zealand eScience Infrastructure (NeSI).
The resulting HMC samples provide a reference posterior approximation for comparison with the SGVB estimates and uncertainty quantification results. The sequential SGVB and HMC procedures described above are collectively referred to as the componentwise approach and are termed componentwise SGVB and componentwise HMC, respectively. For brevity, these componentwise implementations are referred to as SGVB and HMC in the remainder of the main text, unless otherwise stated.

\subsection{Coherence}
In the frequency domain, the strength of association between different components of a multivariate time series can change with frequency. We use coherence to summarize this association at each frequency $f_k$. 
For each spectral density matrix $\S(f_k)$, the off-diagonal element $\S_{ij}(f_k)$ represents the cross-spectrum between components $i$ and $j$, where $i\neq j$, while the diagonal element $\S_{ii}(f_k)$ represents the spectral density of the $i$th component. The coherence between components $i$ and $j$ at frequency $f_k$, denoted by $C_{ij}(f_k)$, is defined as:
\begin{align}\label{coherence}
C_{ij}(f_k) = \frac{|\S_{ij}(f_k)|}{\sqrt{\S_{ii}(f_k)\S_{jj}(f_k)}}.
\end{align}
The value of $C_{ij}(f_k)$ lies between 0 and 1. A value close to 0 indicates little or no linear association between the two components at frequency $f_k$, whereas a value close to 1 indicates a strong linear association.

\section{Simulation study}
\label{sec:simulation}
To evaluate the accuracy and computational efficiency of the proposed SGVB method, we conduct a simulation study using two bivariate stationary time series models: a vector autoregressive model of order 2, VAR(2),
\begin{align*}
\Z_t = \begin{pmatrix}0.5 & 0 \\0 & -0.3\end{pmatrix}\Z_{t-1}+\begin{pmatrix}0 & 0 \\0 & -0.5\end{pmatrix}\Z_{t-2}+\underline{e}_t,
\end{align*}
where
\begin{align*}
\underline{e}_t\overset{iid}{\sim}N_2 \left(\bm{0}, \begin{pmatrix}1 & 0.9 \\0.9 & 1\end{pmatrix} \right),
\end{align*}
and a vector moving-average model of order 1, VMA(1),
\begin{align*}
\Z_t =\underline{e}_t +\begin{pmatrix}-0.75 & 0.5 \\0.5 & 0.75\end{pmatrix}\underline{e}_{t-1}
\end{align*}
where 
\begin{align*}
\underline{e}_t\overset{iid}{\sim}N_2 \left(\bm{0}, \begin{pmatrix}1 & 0.5 \\0.5 & 1\end{pmatrix}\right).
\end{align*}
For each model, 500 independent bivariate time series realizations of length 819,200 are generated. Each realization is then divided into blocks of length $n_b = 256, 512$ and $1{,}024$. The corresponding numbers of blocks are $N_b=3{,}200$, $1{,}600$, and $800$. For each block length, no window function is applied to the blocks, and the spectral density matrix is estimated using both SGVB method and the HMC sampler described in Section~\ref{subsec:sgvb_details}, with the number of basis functions fixed at $M=20$ throughout the simulation study.
For SGVB, the MAP optimization in phase 1 is performed by \texttt{Adam} optimizer with a learning rate of $10^{-3}$ for 10,000 iterations, and the ELBO is maximized in phase 2 with a learning rate of 0.05 for 600 iterations. For HMC, we generate a Markov chain of 80,000 iterations and discard the first 20,000 as burn-in, and the remaining chain is
thinned by retaining every 10th state. These SGVB and HMC settings are kept fixed for all posterior components in every simulation realization.

The resulting SGVB and HMC estimates are compared with the corresponding true spectral density matrix. 
For each realization, we measure the accuracy of the spectral density matrix estimates using the $L_2$ error,

\begin{equation}
 ||\hat{\S} - \S||_{L_2}  \approx \left(\frac{1}{N} \sum_{k=1}^{N}||\hat{\S}(f_k)-\S(f_k)||_F^2 \right)^{1/2}\, ,
\end{equation}
where $\hat{\S}(f_k)$ and $\S(f_k)$ denote the estimated PSD matrix and the true PSD matrix at frequency  $f_k$ , respectively, and $\|\cdot\|_F$ denotes the Frobenius norm.
For a complex-valued $p \times p$ matrix $\mathbf{A}$, the Frobenius norm is defined by
\begin{equation}
\left\|\mathbf{A}\right\|_F
=\left(\sum_{i=1}^{p}\sum_{j=1}^{p}|A_{ij}|^2 \right)^{1/2}.
\end{equation}

\begin{figure}
  \centering
  \includegraphics[width=1\linewidth]{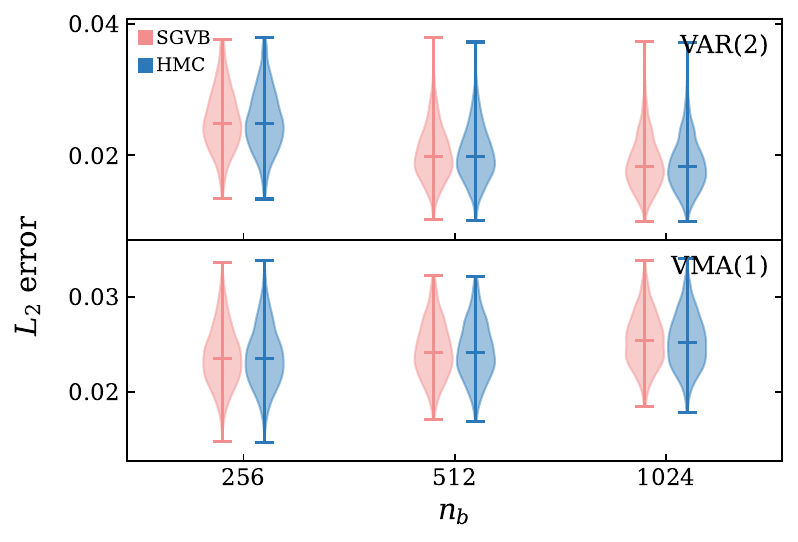}
  \caption{Violin plots of the $L_2$ estimation errors for the SGVB and HMC methods over 500 simulated realizations. Results are shown for the VAR(2) model in the upper panel and the VMA(1) model in the lower panel, under the three block lengths $n_b=256,512$ and $1{,}024$.}
  \label{fig:sim_error_violins}
\end{figure}

\begin{figure}
  \centering
  \includegraphics[width=1\linewidth]{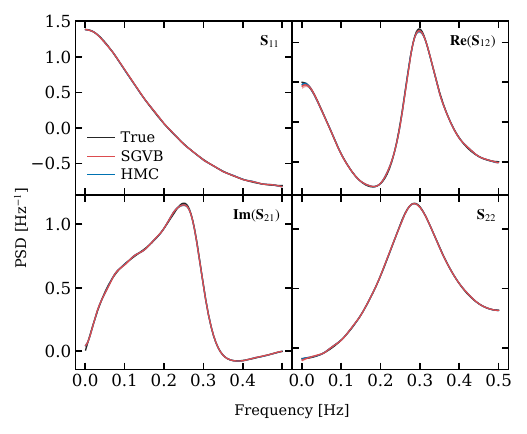}
  \caption{PSD estimates for a VAR(2) realization with block length $n_b=1{,}024$. Black curves denote the true PSD, while red and blue curves denote the posterior median PSD estimates from SGVB and HMC, respectively. Shaded regions show the corresponding pointwise 90\% credible intervals.
  }
  \label{fig:var2_1024}
\end{figure}

Figure~\ref{fig:sim_error_violins} summarizes the distributions of the $L_2$ estimation errors over 500 realizations for the VAR(2) and VMA(1) models under the three block lengths. The upper and lower panels correspond to the VAR(2) and VMA(1) models, respectively.
For the VAR(2) model, the median $L_2$ estimation errors of both SGVB and HMC decrease as the block length increases.
For the VMA(1) model, the median errors increase slightly with block length. Given a fixed total time series length, increasing the block length improves the frequency resolution and can improve the estimation accuracy when the spectrum varies substantially across frequencies. However, it also reduces the number of block periodograms contributing to the summed periodogram matrix in likelihood. The resulting estimates may therefore have greater variability. In the VMA(1) setting, this increased variability may slightly outweigh the potential benefit of improved frequency resolution.
Across both models and all block lengths, the $L_2$ error distributions of SGVB and HMC are very similar. The median  $L_2$ errors from HMC are slightly smaller for some cases, but they remain very close to the corresponding SGVB errors. This indicates that the SGVB approximation achieves estimation accuracy comparable to HMC.
Figure~\ref{fig:var2_1024} shows the SGVB and HMC PSD estimates for a VAR(2) realization with block length $n_b=1{,}024$. Both the SGVB and HMC posterior medians closely follow the true PSD across the frequency range. For most frequencies, the true PSD lies within the corresponding pointwise 90\% credible intervals of both methods.
%As illustrated in Figure~\ref{fig:sim_speed_violins},the computational speed-up factor increases with the block length $n_b$ for both the VAR(2) and VMA(1) models. Across the three block lengths, SGVB requires approximately $1/20$ to $1/40$ of the computational time required by HMC. For the largest block length, $n_b=1{,}024$, the median speed-up factor is close to 40 for both models. Together with the close agreement between the $L_2$ error distributions of SGVB and HMC, these results show that SGVB provides point estimates with accuracy comparable to HMC at substantially lower computational cost.

Table~\ref{table:simstudy} further demonstrates the quantification of the uncertainty of SGVB and HMC through the probabilities of pointwise coverage and the widths of credible intervals. HMC tends to provide higher pointwise coverage than SGVB. This behavior can be explained by the mean field approximation~\cite{Blei2017} and the diagonal Gaussian surrogate distribution, which does not explicitly capture posterior dependence among parameters. In addition, minimizing the reverse KL divergence tends to penalize assigning probability mass to regions where the posterior density is low. This may encourage the surrogate distribution to concentrate around a high density region of the posterior. These features may cause the variational approximation to underestimate posterior uncertainty, which can result in narrower credible intervals and lower pointwise coverage~\cite{Blei2006,Wang2005}.
Potential remedies include using more flexible variational families or alternative divergence measures\cite{Kingma2013, Rezende2015,Yang2020,Higgins2017}. However, these strategies typically introduce additional computational cost. Since the main objective of this work is scalable spectral density estimation for long multivariate time series, we retain the SGVB approximation as a computationally efficient compromise.
HMC pointwise coverage also remains below the nominal 90\%, ranging from approximately 63\% to 80\%. A possible reason for this undercoverage is the funnel geometry illustrated by Neal's funnel. Such geometry can arise under the centered hierarchical shrinkage parameterization~\citep{BetancourtGirolami2015}. When the shrinkage scales are small, the conditional distribution of the basis coefficients becomes sharply concentrated, producing strong posterior curvature that can impede HMC exploration. Investigation of non-centered parameterizations is left to future work.
SGVB requires substantially less computation time than HMC for both simulation models and all block lengths. Although the computation times of both methods increase with block length, SGVB retains this substantial computational advantage throughout. Together with the close agreement between the median $L_2$ errors of SGVB and HMC, these results show that SGVB provides point estimates with accuracy comparable to HMC at substantially lower computational cost.

\input{table}

To assess and validate the HMC settings used throughout the simulation study, four independent chains were run for a representative VAR(2) realization with block length $n_b=1{,}024$. For the basis coefficients determining the spectral density estimates, the median split potential scale reduction factors ($\hat R$) were close to 1, and all corresponding \ac{ESS} values exceeded 4,500. These diagnostics supported applying the same HMC settings to all posterior components and simulation realizations.
In contrast, the hierarchical prior hyperparameters showed poorer mixing, likely reflecting weaker identification and stronger posterior dependence within the shrinkage hierarchy. However, the induced spectral density estimates were insensitive to this poorer mixing.

\section{Applications to LISA}\label{sec:applications}

The proposed SGVB method is applied to two one-year simulated LISA instrumental-noise datasets: the symmetric \texttt{noise4a} realization \citep{lisa_noise4a} and the asymmetric \texttt{noise5a} realization \citep{lisa_noise5a}. Both datasets are represented in the second-generation TDI Michelson channels $X$, $Y$, and $Z$ and were generated from the LISA Data Challenge Spritz noise model \citep{ldcspritz}. They assume a fixed inter-spacecraft light-travel time of 8.3 seconds, so orbital breathing and time-dependent arm-length modulation are not included. The two datasets therefore provide controlled stationary noise examples for assessing multivariate spectral density estimation, while differing in the degree of instrumental-noise asymmetry.

The analysis uses the same eigenbasis representation of the blocked multivariate Whittle likelihood as \citet{Vajpeyi2026}, but differs in the spectral representation and posterior inference. \citet{Vajpeyi2026} model the Cholesky components using penalized B-splines with hierarchical Gaussian smoothing priors and use SVI only to initialize NUTS. In contrast, the present work uses cosine basis expansions with a discounted regularized horseshoe prior and employs SGVB as the primary posterior approximation. For both \texttt{noise4a} and \texttt{noise5a}, HMC is also implemented as a sampling-based benchmark, allowing the SGVB approximation to be compared with posterior samples from the same model.

The same SGVB and HMC settings are used for both LISA analyses. For each posterior component, SGVB uses the \texttt{Adam} optimizer to perform the phase 1 MAP optimization with a learning rate of $3\times10^{-4}$ for 10,000 iterations, followed by the phase 2 ELBO optimization with a learning rate of $0.05$ for 600 iterations. For HMC, a Markov chain of 80,000 iterations is generated for each posterior component. The first 20,000 iterations are discarded as burn-in, and the remaining chain is thinned by retaining every 20th
state.

Both datasets are processed using the same blockwise scheme. Each one-year time series is divided into non-overlapping blocks of $n_b=16{,}384$ samples. Within each block, the sample mean is removed and a Kaiser window with shape parameter $\beta=30$ is applied before constructing the likelihood. Because the same taper and block length are used for the \texttt{noise4a} and \texttt{noise5a} analyses, the corresponding normalized equivalent noise bandwidth is $N_{\rm bw}=3.12$ in both cases. This relatively large value of $\beta$ is used to reduce spectral leakage caused by the steep low-frequency behavior of the LISA TDI spectra and by the sharp transfer-function features within the analysis band.

The \texttt{noise4a} dataset is the symmetric configuration. The \acs{OMS} and \acs{TM} noise components are assigned identical baseline noise levels across the relevant optical paths, giving a comparatively idealized instrumental-noise realization. The dataset contains 6{,}291{,}456 samples per TDI channel and is sampled at 0.2 Hz. With $n_b=16{,}384$, this gives 384 non-overlapping blocks, each with duration $T_b=81{,}920$ seconds. The fundamental block frequency is therefore $1/T_b\approx1.22\times10^{-5}\,\mathrm{Hz}$, and the Nyquist frequency is 0.1 Hz. We estimate the spectral density matrix over $10^{-4}$--$10^{-1}\,\mathrm{Hz}$, retaining 8{,}184 Fourier bins from each block to construct the blocked likelihood. The Cholesky components are represented using $M=250$ basis functions, selected through a low-frequency sensitivity analysis based on the agreement between the SGVB posterior median PSD and the Welch estimate. Further details are provided in Appendix~\ref{app:basis_noise4a}.

\begin{figure}
  \centering
  \includegraphics[width=\columnwidth]{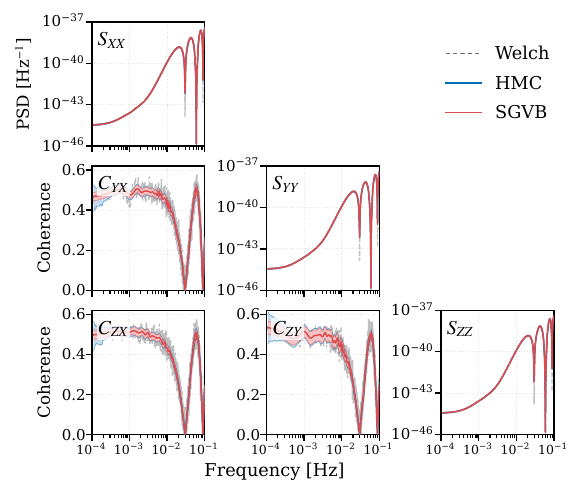}
  \caption{Estimated PSDs and pairwise coherences for the \texttt{noise4a} dataset in a lower-triangular layout. The diagonal panels show the PSDs of the TDI Michelson channels $X$, $Y$, and $Z$, while the lower-triangular panels show the corresponding pairwise coherences. The dashed gray curves denote the Welch estimates. The blue and red curves denote the HMC and SGVB posterior medians, respectively. Shading shows the corresponding 90\% uniform credible bands.}
  \label{fig:noise4a}
\end{figure}

The \texttt{noise5a} dataset uses the same fixed-delay TDI Michelson construction, but introduces asymmetric instrumental noise. The \acs{OMS} and \acs{TM} noise amplitudes associated with the six movable optical sub-assemblies (MOSAs) are independently scaled around their baseline values, with scaling factors drawn from $\mathcal{U}(0.5,\,2.0)$. This breaks the symmetry of \texttt{noise4a} and produces a more heterogeneous noise configuration. The dataset contains 15{,}777{,}792 samples per TDI channel and is sampled at 0.5 Hz. With the same block length $n_b=16{,}384$, the analysis uses 963 non-overlapping blocks, each with duration $T_b=32{,}768$ seconds. The corresponding fundamental block frequency is $1/T_b\approx3.05\times10^{-5}\,\mathrm{Hz}$, and the Nyquist frequency is 0.25 Hz. For this dataset, the likelihood is constructed from the 3{,}273 Fourier bins per block lying in $10^{-4}$--$10^{-1}\,\mathrm{Hz}$, and the Cholesky components are represented using $M=400$ basis functions, selected through the same low-frequency sensitivity analysis. Further details are provided in Appendix~\ref{app:basis_noise5a}.
%\textcolor{blue}{Unlike the P-spline analysis of \citet{Vajpeyi2026}, which coarse-grains adjacent Fourier bins to $N_c=1{,}024$ and uses NUTS sampling, SGVB retains 8{,}184 and 3{,}273 Fourier bins per block within the selected frequency range for the \texttt{noise4a} and \texttt{noise5a} datasets, respectively. The computational times reported in the two studies are therefore not directly comparable. Nevertheless, the present results demonstrate that SGVB remains computationally feasible despite the larger number of Fourier bins are used.}

\begin{figure}
  \centering
  \includegraphics[width=\columnwidth]{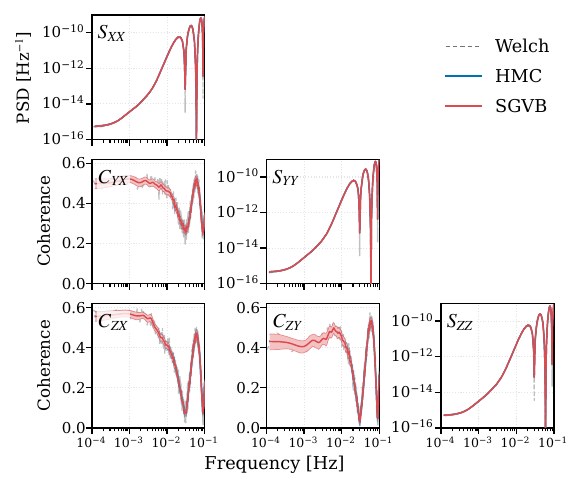}
  \caption{Estimated PSDs and pairwise coherences for the \texttt{noise5a} dataset in a lower-triangular layout. The diagonal panels show the PSDs of the TDI Michelson channels $X$, $Y$, and $Z$, while the lower-triangular panels show the corresponding pairwise coherences. The dashed gray curves denote the Welch estimates. The blue and red curves denote the HMC and SGVB posterior medians, respectively. Shading shows the corresponding 90\% uniform credible bands.}
  \label{fig:noise5a}
\end{figure}

\input{table2}
Figures~\ref{fig:noise4a} and \ref{fig:noise5a} illustrate the estimated PSDs for the three TDI Michelson channels $X, Y$, and $Z$, together with the corresponding pairwise coherence estimates over the analysis band $10^{-4}-10^{-1}$ Hz, for the $\texttt{noise4a}$ and $\texttt{noise5a}$ datasets, respectively. The 90\% uniform credible bands summarize the posterior uncertainty in the SGVB and HMC estimates. In both cases, the SGVB PSD estimates closely follow both the HMC PSD estimates and the Welch estimates across the analysis band. The pairwise coherences estimated by both SGVB and HMC also show close agreement with the Welch-based coherences.
For each dataset, the corresponding analytic reference PSD $\S_{\mathrm{ref}}(f)$ is used for quantitative comparison. It is constructed by propagating the link-level \acs{OMS} and \acs{TM} noise spectra specified by the LDC Spritz noise model, including the dataset-specific \acs{MOSA} amplitude scalings, through the fixed-delay, second-generation Michelson TDI transfer functions \citep{Nam2023,lisa_noise4a}. The reference coherence used below is computed from this analytic reference PSD using Eq.~\eqref{coherence}.
The PSD curves exhibit sharp dips, particularly in the range $10^{-2}$--$10^{-1}\,\mathrm{Hz}$. These dips arise from finite-arm-length TDI transfer functions, which strongly attenuate the spectral power at particular frequencies \citep{TintoMassimo2021Ti}. Consequently, the PSD and coherence estimates exhibit pronounced local fluctuations near these frequencies. To avoid this influence in the quantitative comparisons, the frequency intervals centered at 0.03, 0.06, and 0.090 Hz, each with a half-width of 1 mHz, are removed before evaluating the error metrics below.
Let $\mathcal{F}$ denote the retained frequencies after these removals. To quantify the agreement between each posterior median PSD estimate and the corresponding analytic reference PSD, the \ac{RIAE} is calculated as:
\begin{align}\label{RIAE}
\mathrm{RIAE} = 
\frac{\int_{\mathcal{F}}
\left\|
\hat{\mathbf{S}}(f) -
\mathbf{S}_{\mathrm{ref}}(f)
\right\|_F \, \mathrm{d}f}
{
\int_{\mathcal{F}}
\left\|
\mathbf{S}_{\mathrm{ref}}(f)
\right\|_F \, \mathrm{d}f
},
\end{align}
where $\hat{\mathbf{S}}(f)$ denotes the posterior median PSD estimate obtained by SGVB or HMC. 
Using the same retained frequency set $\mathcal{F}$, the accuracy of the corresponding pairwise coherence estimates is further assessed
using the \ac{MAE},
\begin{align}
\mathrm{MAE}(C_{ij}) =
\frac{1}{N_{\mathcal{F}}}
\sum_{f_k\in\mathcal{F}}
\left\lvert
\hat{C}_{ij}(f_k)
-
C_{ij,\mathrm{ref}}(f_k)
\right\rvert ,
\label{coherence_mae}
\end{align}
where $N_{\mathcal{F}}$ denotes the number of retained Fourier frequency points in $\mathcal{F}$, $ij\in\{XY,YZ,ZX\}$, $\hat{C}_{ij}(f_k)$ denotes the estimated coherence, and $C_{ij,\mathrm{ref}}(f_k)$ denotes the reference coherence.
Table~\ref{tab:lisa_results} summarizes the accuracy and computational time of SGVB and HMC for the two LISA noise configurations. Across both datasets, the two methods yield comparable RIAE values and closely matched \acs{MAE} values for all three pairwise coherence estimates. These results indicate that, for both the symmetric and asymmetric noise configurations, SGVB produces PSD and coherence estimates consistent with the HMC benchmark. Despite this agreement in estimation accuracy, SGVB is approximately $37$ times faster than HMC for \texttt{noise4a} and $27$ times faster for \texttt{noise5a} under the same posterior model.

These findings are consistent with the broader conclusions of the related LISA analyses, while addressing a different inferential objective. For the same \texttt{noise4a} and \texttt{noise5a} simulations, \citet{Vajpeyi2026} find that a full multivariate P-spline model recovers the spectral matrix in both cases, whereas restricting the noise covariance to be diagonal in the $A,E,T$ basis substantially degrades accuracy for the asymmetric \texttt{noise5a} configuration. Their unrestricted model reports one-year \acs{RISE} values of $5.0\times10^{-3}$ and $8.6\times10^{-4}$ for \texttt{noise4a} and \texttt{noise5a}, respectively, with SVI--NUTS runtimes of 853 and 1,089 seconds. \citet{Liu2026} analyze \texttt{noise4a} using their matrix-gamma process model and likewise recover the PSD and coherence structure, with slight underestimation at low frequencies. They do not report a scalar LISA error or an end-to-end LISA runtime. The present analysis complements these studies by testing whether SGVB can replace sampling as the production inference method. Its \acs{RIAE} values are $2.667\times10^{-2}$ and $3.37\times10^{-2}$, with runtimes of 473 and 310 seconds, for \texttt{noise4a} and \texttt{noise5a}, respectively, while the same-model HMC fits provide a controlled benchmark.

The numerical error values and wall-clock times across the three studies are not directly comparable. In particular, RIAE and RISE are different loss functions, and the analyses use different priors, basis dimensions, coarse-graining, tapering and tempering choices, frequency conditioning, implementations, and computing environments. For example, \citet{Vajpeyi2026} use 1,024 coarse frequency bins and approximately week-long blocks, the present work retains 8,184 and 3,273 frequency bins using blocks of 16,384 samples, and \citet{Liu2026} use approximately day-long, Hann-tapered blocks for \texttt{noise4a}.

As an additional implementation check, we also fitted the LISA datasets using a joint SGVB approximation to the full posterior based on the taper-adjusted blocked eigenbasis likelihood in Eqs.~\eqref{eigenbasis_likelihood} and~\eqref{nbw}, rather than applying SGVB sequentially to the componentwise posterior factors in Eq.~\eqref{eq:component_posterior}. The joint SGVB optimizes a single variational approximation over all basis coefficients and shrinkage parameters simultaneously. For both the \texttt{noise4a} and \texttt{noise5a} datasets, the posterior median PSD estimates from the joint SGVB are very close to those obtained from the componentwise SGVB approach. However, the joint SGVB required 17.5 minutes for \texttt{noise4a} and 6.4 minutes for \texttt{noise5a}, compared with 7.9 minutes and 5.2 minutes for the componentwise SGVB, respectively. This comparison indicates that the componentwise SGVB approach gives similar spectral estimates while reducing the computational time.

\section{Discussion}\label{sec:discussion}
In this work, we presented a componentwise variational Bayesian framework for estimating spectral density matrices of long multivariate stationary time series. The method combines a blocked multivariate Whittle likelihood with the eigenbasis representation of the summed periodogram matrices, and parameterizes the inverse spectral density matrix through Cholesky components represented by basis functions. SGVB is used as the primary posterior approximation and is applied sequentially to the posterior components, optimizing a tractable surrogate distribution for each component rather than sampling directly from the posterior.

The simulation study based on the VAR(2) and VMA(1) models shows that SGVB gives posterior median spectral estimates close to those obtained from HMC, while requiring substantially less computation time. This supports the use of SGVB as a scalable substitute for HMC when the main target is fast posterior-median spectral estimation. At the same time, the SGVB credible intervals are generally narrower than the corresponding HMC intervals. This reflects the more concentrated variational approximation and should be taken into account when using SGVB interval estimates \cite{Kingma2013, Rezende2015,Yang2020,Higgins2017}.

The applications to the \texttt{noise4a} and \texttt{noise5a} datasets show that the method performs well for controlled LISA instrumental-noise examples. Across both the symmetric and asymmetric noise configurations, the SGVB posterior median estimates closely agree with the HMC and Welch estimates. The comparable RIAE values with respect to the analytic reference PSD, together with the MAE values for the coherence estimates, indicate that SGVB accurately captures the multivariate spectral structure in both datasets. SGVB also requires substantially less computation time than HMC for these year-long LISA data.

A limitation of the present LISA application is that the analysis assumes stationarity over the full one-year dataset. The \texttt{noise4a} and \texttt{noise5a} simulations are useful controlled test cases, but real LISA data will contain gaps, glitches, slowly drifting noise levels, and time-dependent arm lengths due to orbital breathing. The blockwise structure of the likelihood provides a natural starting point for future extensions that exclude or down-weight problematic blocks, allow piecewise-stationary spectra or estimate time-evolving spectral density matrices across shorter time intervals, in the spirit of Bayesian time-frequency methods for multivariate spectra \citep{LiKrafty2019}.

Another direction for improving scalability is computational rather than statistical. The current implementation fits the componentwise SGVB posterior approximations sequentially and reports timings from a single CPU core. Since the Cholesky factorization separates the posterior into componentwise terms, future implementations could optimize the SGVB surrogate for multiple posterior components in parallel and then combine the resulting surrogate draws to form samples of the full spectral density matrix. In addition, the MAP and ELBO optimization steps are TensorFlow gradient computations over large frequency grids, so vectorized multi-core or GPU implementations could further reduce wall-clock time without changing the underlying posterior model.

A further natural direction is to extend the framework from noise-only spectral estimation to joint signal-and-noise inference. In this setting, the spectral density model could be combined with a signal model so that the instrumental noise spectrum and signal parameters are inferred simultaneously. This would be particularly relevant for stochastic gravitational-wave background analyses, where uncertainty in the instrumental noise spectrum can directly affect inference on weak astrophysical signals. Further work is needed to assess identifiability between the noise and signal components in such joint models.

\section*{Data and Software Availability}
The software developed for this research will be available on GitHub as version 2.0 of the \texttt{sgvb\_psd} Python package~\citep{NZ_Gravity_sgvb_psd_2024}.
Documentation and examples of the software can be found at \url{https://nz-gravity.github.io/sgvb_psd/}. 
The data associated with this work will be made available by the authors upon reasonable request.

\begin{acknowledgments}
We thank Jean-Baptiste Bayle for producing and providing the simulated LISA \texttt{noise4a} and \texttt{noise5a} datasets used in this work. The authors gratefully acknowledge support from the Marsden Fund Council grant MFP-UOA2131, funded by the New Zealand Government and managed by the Royal Society Te Ap\={a}rangi. We thank the New Zealand eScience Infrastructure (NeSI) for the use of its high-performance computing facilities and the Centre for eResearch at the University of Auckland for technical support. We also thank Yixuan Liu for assistance with the simulation study.
\end{acknowledgments}

\appendix

%\avi{Add appendix comparing impact of factorised non factorised versions -- timing, etc should be much faster. Even for LISA is fine.}

\section{Priors}
\label{priors}
To avoid shrinking the intercept and linear term in the expansion of the diagonal Cholesky terms $\log \delta_{jk}^2$ to zero, weakly informative Gaussian priors $\gamma_{j,0}$ and $\gamma_{j,1}$ are assigned as $\mathcal{N} \left(0, 10\right)$ for $j=1,\dots,p$. For the remaining diagonal coefficients,
\begin{equation}
\gamma_{j,s}|\lambda_{js},\tau_j \sim \mathcal{N} \left(0, \frac{c^2\tau_j^2\lambda_{js}^2}{c^2+\tau_j^2\lambda_{js}^2}\right).
\end{equation}
Similarly, for the real and imaginary parts of the off-diagonal Cholesky parameters $\theta_{jl}^{(k)}$, 
\begin{equation}
\alpha_{jl,s}|\lambda_{jls,(re)},\tau_{jl} \sim \mathcal{N} \left(0, \frac{c^2\tau_{jl}^2\lambda_{jls,(re)}^2}{c^2+\tau_{jl}^2\lambda_{jls,(re)}^2} \right)
\end{equation}
and 
\begin{equation}
\beta_{jl,s}|\lambda_{jls,(im)},\tau_{jl} \sim \mathcal{N} \left(0, \frac{c^2\tau_{jl}^2\lambda_{jls,(im)}^2}{c^2+\tau_{jl}^2\lambda_{jls,(im)}^2} \right).
\end{equation}
The global shrinkage parameters are assigned as Half-Cauchy distributions, i.e., $\tau_j, \tau_{jl} \sim C^+(0,c_\tau)$. The hyperparameters are fixed at $c=2$ and $c_{\tau}=0.01$. The local shrinkage parameters $\lambda_{js}$, $\lambda_{jls,(re)}$, and $\lambda_{jls,(im)}$ are assigned as Half-Cauchy distributions with scale $c_s$, i.e., $C^+(0,c_s)$, where $c_s$ is defined through a decreasing sigmoid discount function, i.e.,
\begin{equation}
c_s=\text{Sig}(-as+b)=\frac{1}{1+\e^{as-b}},
\end{equation}
where $a$ determines the rate of decreases of $c_s$ with respect to $s$, while $b$ controls the location of this decrease, the values are set to $a=1$ and $b=M$. The scale $c_s$ decreases monotonically as the basis functions used increases. This discounting mechanism imposes stronger shrinkage on higher order basis coefficients, regularizing the basis expansion while still allowing coefficients supported by the data to remain active.

% \section{SGVB--HMC coherence differences}
% \label{app:coherence_difference}
\section{Basis-function selection for the LISA noise simulations}
\label{app:basis_selection}
The number of basis functions $M$ controls the flexibility of the SGVB spectral density estimate. If $M$ is too small, the resulting estimate can be too smooth to capture local spectral features. If $M$ is too large, the additional high-order basis functions can introduce unnecessary local variation. For the LISA applications, the cosine basis functions are evaluated on the normalized linear frequency scale used in the basis expansion. Over the three-decade band $10^{-4}$--$10^{-1}\,\mathrm{Hz}$, the lowest decade occupies only a small fraction of this scale. Consequently, discrepancies confined to this low-frequency region contribute little to the overall likelihood. When $M$ is large, the basis functions can therefore introduce local variation in the low-frequency PSD that is weakly constrained by the data, making low-frequency estimates especially sensitive to $M$.
For both \texttt{noise4a} and \texttt{noise5a}, we therefore use an empirical low-frequency diagnostic. We fit the SGVB model over a grid of candidate values of $M$ and compare the SGVB posterior median spectral density with the corresponding Welch estimate over $10^{-4}$--$10^{-3}\,\mathrm{Hz}$. The discrepancy is measured using the RIAE defined in Eq.~\eqref{RIAE}, with $\S_{\mathrm{ref}}(f)$ replaced by $\S_{\mathrm{Welch}}(f)$ and the integration restricted to this low-frequency band. The final value of $M$ is chosen near the beginning of the plateau in the RIAE curve, balancing low-frequency fit against unnecessary model complexity.

To assess the accuracy of the SGVB posterior median PSD estimate obtained for each candidate value of $M$, we compare it with $\S_{\mathrm{ref}}(f)$ using the frequency-resolved relative absolute error (RAE), which is the pointwise counterpart of the RIAE in Eq.~\eqref{RIAE}:
\begin{align}
\mathrm{RAE}(f)
=
\frac{\left\|\hat{\S}_M(f)-\S_{\mathrm{ref}}(f)\right\|_F}
{\left\|\S_{\mathrm{ref}}(f)\right\|_F},
\qquad f\in\mathcal{F},
\label{eq:rae}
\end{align}
where $\hat{\S}_M(f)$ denotes the SGVB posterior median spectral density estimate obtained using $M$ basis functions. This diagnostic is evaluated for the representative values of $M$ shown in the RAE panels and is used to check, rather than determine, the Welch-based selections of $M$. The dip frequencies and their surrounding neighborhoods are excluded.

\subsection{Selection for \texttt{noise4a}}
\label{app:basis_noise4a}
\begin{figure}[t]
  \centering
  \subfloat[PSD and coherence estimates.\label{fig:noise4a_basis_psd}]{%
    \includegraphics[width=0.98\columnwidth]{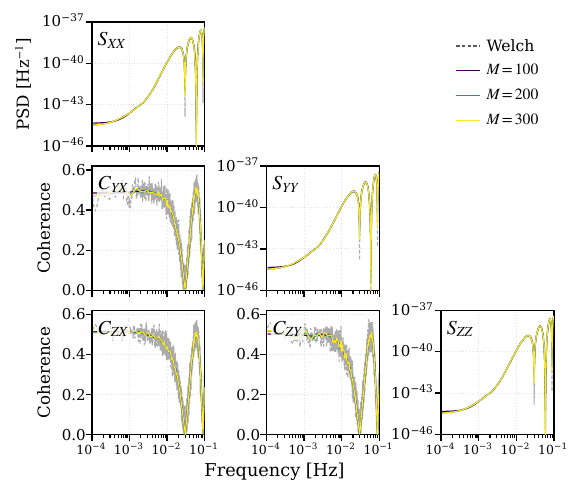}}
  \par\vspace{-0.3em}
  \subfloat[Low-frequency RIAE.\label{fig:noise4a_riae}]{%
    \includegraphics[width=0.49\columnwidth]{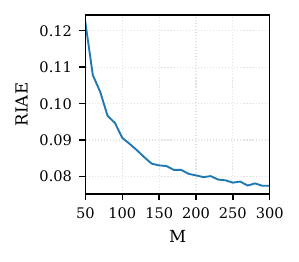}}%
  \hfill
  \subfloat[$\mathrm{RAE}(f)$.\label{fig:noise4a_basis_relative_error}]{%
    \includegraphics[width=0.49\columnwidth]{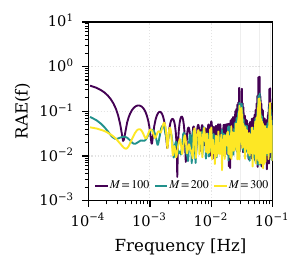}}
  \caption{Basis-function sensitivity analysis for \texttt{noise4a}. (a) Welch estimate and SGVB posterior medians for $M=100$, $200$, and $300$; the diagonal panels show the channel PSDs and the lower-triangular panels show pairwise coherences. (b) Low-frequency RIAE for $M=50,60,\ldots,300$. (c) RAE relative to $\S_{\mathrm{ref}}(f)$ for $M=100$, $200$, and $300$. The dip frequencies and their surrounding neighborhoods are omitted from panel (c).}
  \label{fig:noise4a_basis}
\end{figure}

For the \texttt{noise4a} dataset, we first examine SGVB fits over a broad range of values of $M$. Figure~\ref{fig:noise4a_basis_psd} displays the Welch estimate together with SGVB posterior median estimates for $M=100$, $200$, and $300$. This comparison shows that the low-frequency SGVB estimates are sensitive to the basis size and motivates the quantitative RIAE assessment.

We then compute the RIAE between the SGVB posterior median estimates and the Welch estimate for the candidate values $M=50,60,\ldots,300$.
As shown in Figure~\ref{fig:noise4a_riae}, the RIAE decreases rapidly for small values of $M$ and begins to level off around $M=250$. Although the RIAE continues to decrease slightly for larger values of $M$ within the retained candidate set, the additional improvement beyond this point is limited. 
We therefore set $M=250$ for the \texttt{noise4a} analysis.

As shown in Figure~\ref{fig:noise4a_basis_relative_error}, the frequency-dependent fluctuations in $\mathrm{RAE}(f)$ generally decrease as $M$ increases. The curves for $M=200$ and $M=300$ are broadly comparable across the analysis band. Together with the RIAE plateau around $M=250$, this supports the selection of the intermediate value $M=250$.

\subsection{Selection for \texttt{noise5a}}
\label{app:basis_noise5a}
\begin{figure}[t]
  \centering
  \subfloat[PSD and coherence estimates.\label{fig:noise5a_basis_psd}]{%
    \includegraphics[width=0.98\columnwidth]{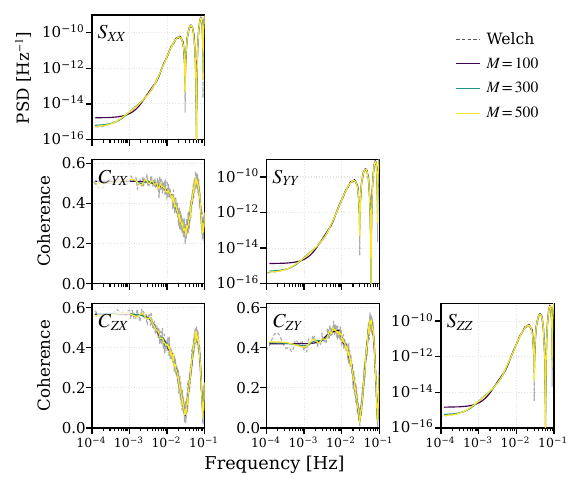}}
  \par\vspace{-0.3em}
  \subfloat[Low-frequency RIAE.\label{fig:noise5a_riae}]{%
    \includegraphics[width=0.49\columnwidth]{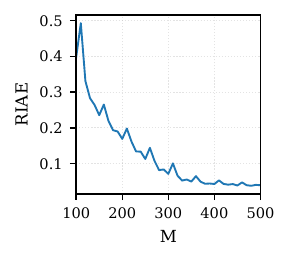}}%
  \hfill
  \subfloat[$\mathrm{RAE}(f)$.\label{fig:noise5a_basis_relative_error}]{%
    \includegraphics[width=0.49\columnwidth]{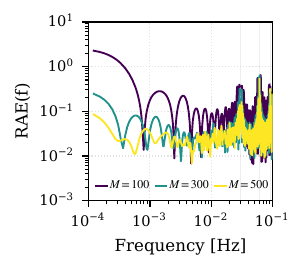}}
  \caption{Basis-function sensitivity analysis for \texttt{noise5a}. (a) Welch estimate and SGVB posterior medians for $M=100$, $300$, and $500$; the diagonal panels show the channel PSDs and the lower-triangular panels show pairwise coherences. (b) Low-frequency RIAE for $M=100,110,\ldots,500$. (c) RAE relative to $\S_{\mathrm{ref}}(f)$ for $M=100$, $300$, and $500$. The dip frequencies and their surrounding neighborhoods are omitted from panel (c).}
  \label{fig:noise5a_basis}
\end{figure}

For the \texttt{noise5a} dataset, we select the number of basis functions $M$ using the same RIAE criterion described above. Figure~\ref{fig:noise5a_basis_psd} shows representative SGVB posterior median estimates for $M=100$, $300$, and $500$, together with the corresponding Welch estimate. In this case, increasing $M$ mainly reduces the low-frequency offset in the PSD estimates, and the curves with larger values of $M$ align more closely with the Welch estimate over the low-frequency band.

The RIAE is calculated for $M=100,\ldots,500$. As shown in Figure~\ref{fig:noise5a_riae}, it decreases rapidly for small and moderate values of $M$ and begins to level off around $M=400$. Increasing the basis size to $M=500$ provides little additional improvement in the RIAE. We therefore set $M=400$ for the \texttt{noise5a} analysis.

Figure~\ref{fig:noise5a_basis_relative_error} shows that $\mathrm{RAE}(f)$ generally decreases and becomes less variable as $M$ increases from 100 to 500. Together with the RIAE plateau around $M=400$, this supports selecting $M=400$ without using the larger basis size.

\bibliography{reference}

\end{document}

%% file: authors.tex
\newcommand{\UoAStats}{Department of Statistics, University of Auckland, 38 Princes St, Auckland, New Zealand}

\newcommand{\AUTMaths}{Department of Mathematical Sciences, Auckland University of Technology, Auckland, New Zealand}

\author{Jianan Liu}
\affiliation{\UoAStats}

\author{Avi Vajpeyi}
\affiliation{\UoAStats}

\author{Renate Meyer}
\affiliation{\UoAStats}

\author{Jeung Eun Lee}
\affiliation{\UoAStats}

\author{Patricio Maturana-Russel}
\affiliation{\UoAStats}
\affiliation{\AUTMaths}

%% file: table.tex
\begingroup
\begin{table*}[!htbp]
\centering
\small
\renewcommand{\arraystretch}{1.5}
\setlength{\tabcolsep}{3pt}

\newcommand{\medmad}[2]{\ensuremath{#1\pm#2}}

\begin{NiceTabular}{l l c *{3}{|cc}}[colortbl-like, cell-space-limits=1pt]
\CodeBefore
  \rowcolors{3}{gray!10}{white}
  \columncolor{white}{1-3}
\Body
\Block{1-9} \\
  &&& \Block{1-2}{\bfseries $n_b = 256$}
     && \Block{1-2}{\bfseries $n_b = 512$}
     && \Block{1-2}{\bfseries $n_b = 1{,}024$} & \\
  & Metric & Scale & SGVB & HMC & SGVB & HMC & SGVB & HMC \\
\hline
\Block{7-1}{\bfseries VAR(2)}
  & $L_2$ error & $10^{-2}$
  & \medmad{2.49}{0.31} & \medmad{2.49}{0.31}
  & \medmad{1.98}{0.26} & \medmad{1.98}{0.26}
  & \medmad{1.83}{0.25} & \medmad{1.83}{0.25} \\
  & Pointwise coverage & \%
  & \medmad{61.4}{4.1} & \medmad{62.6}{4.3}
  & \medmad{73.6}{4.9} & \medmad{75.8}{4.9}
  & \medmad{77.0}{5.3} & \medmad{79.8}{5.6} \\
  & $\S_{11}$ CI width & $10^{-2}$
  & \medmad{1.22}{0.05} & \medmad{1.08}{0.05}
  & \medmad{1.22}{0.04} & \medmad{1.08}{0.04}
  & \medmad{1.21}{0.04} & \medmad{1.08}{0.04} \\
  & $\Re \S_{12}$ CI width & $10^{-2}$
  & \medmad{0.95}{0.04} & \medmad{0.96}{0.04}
  & \medmad{1.04}{0.11} & \medmad{0.98}{0.05}
  & \medmad{1.07}{0.09} & \medmad{1.03}{0.07} \\
  & $\Im \S_{12}$ CI width & $10^{-2}$
  & \medmad{1.63}{0.04} & \medmad{1.00}{0.02}
  & \medmad{1.60}{0.06} & \medmad{1.00}{0.03}
  & \medmad{1.59}{0.05} & \medmad{0.99}{0.03} \\
  & $\S_{22}$ CI width & $10^{-2}$
  & \medmad{1.25}{0.04} & \medmad{1.51}{0.04}
  & \medmad{1.26}{0.04} & \medmad{1.52}{0.04}
  & \medmad{1.27}{0.06} & \medmad{1.52}{0.05} \\
  & Time & $-$
  & $31\pm2\,\mathrm{s}$ & $1{,}084\pm66\,\mathrm{s}$
  & $37\pm4\,\mathrm{s}$ & $1{,}147\pm89\,\mathrm{s}$
  & $47\pm3\,\mathrm{s}$ & $1{,}314\pm117\,\mathrm{s}$ \\
\hline
\Block{7-1}{\bfseries VMA(1)}
  & $L_2$ error & $10^{-2}$
  & \medmad{2.35}{0.21} & \medmad{2.35}{0.21}
  & \medmad{2.42}{0.20} & \medmad{2.41}{0.21}
  & \medmad{2.54}{0.20} & \medmad{2.52}{0.21} \\
  & Pointwise coverage & \%
  & \medmad{61.2}{3.4} & \medmad{71.5}{3.5}
  & \medmad{63.4}{3.3} & \medmad{74.8}{3.3}
  & \medmad{62.8}{3.5} & \medmad{75.4}{3.7} \\
  & $\S_{11}$ CI width & $10^{-2}$
  & \medmad{1.94}{0.01} & \medmad{1.66}{0.03}
  & \medmad{1.93}{0.01} & \medmad{1.66}{0.03}
  & \medmad{1.93}{0.01} & \medmad{1.66}{0.03} \\
  & $\Re \S_{12}$ CI width & $10^{-2}$
  & \medmad{0.81}{0.02} & \medmad{1.94}{0.09}
  & \medmad{0.77}{0.02} & \medmad{2.02}{0.08}
  & \medmad{0.76}{0.02} & \medmad{2.06}{0.09} \\
  & $\Im \S_{12}$ CI width & $10^{-2}$
  & \medmad{1.17}{0.27} & \medmad{1.88}{0.08}
  & \medmad{1.05}{0.17} & \medmad{1.88}{0.07}
  & \medmad{1.06}{0.07} & \medmad{1.89}{0.06} \\
  & $\S_{22}$ CI width & $10^{-2}$
  & \medmad{4.26}{0.03} & \medmad{4.95}{0.05}
  & \medmad{4.20}{0.03} & \medmad{4.96}{0.05}
  & \medmad{4.17}{0.03} & \medmad{4.96}{0.04} \\
  & Time & $-$
  & $29\pm1\,\mathrm{s}$ & $1{,}043\pm82\,\mathrm{s}$
  & $34\pm1\,\mathrm{s}$ & $1{,}080\pm65\,\mathrm{s}$
  & $45\pm3\,\mathrm{s}$ & $1{,}268\pm71\,\mathrm{s}$ \\
\end{NiceTabular}

\caption{Comparison of the medians and median absolute deviations (MADs) of the $L_2$ error, pointwise coverage, widths of pointwise 90\% credible intervals, and computation time (in seconds), calculated across 500 simulated realizations with a fixed total time-series length of $n=819{,}200$. The Scale column indicates the factor applied to the numerical entries; \% denotes percentages and $-$ indicates no rescaling. Results are shown for SGVB and HMC under block lengths $n_b=256$, $512$, and $1{,}024$ for the VAR(2) and VMA(1) models.}
\label{table:simstudy}
\end{table*}
\endgroup

%% file: table2.tex
% \begin{table*}[t]
% \begin{ruledtabular}
% \begin{tabular}{lcccc}
%  & \multicolumn{2}{c}{$\texttt{noise4a}$}
%  & \multicolumn{2}{c}{$\texttt{noise5a}$} \\
% \cline{2-3} \cline{4-5}
% Metric & SGVB & HMC & SGVB & HMC \\
% \hline
% RIAE
% & 0.02667 & 0.02662
% & 0.0337  & 0.0333 \\
% Time [s]
% & $473$ ($7.9$ min) & $17{,}730$ ($4.93$ h)
% & $310$ ($5.2$ min) & $8{,}256$ ($2.29$ h) \\
% $\mathrm{MAE}_{XY}$
% & 0.00609 & 0.00614
% & 0.00531 & 0.00538 \\
% $\mathrm{MAE}_{YZ}$
% & 0.00580 & 0.00561
% & 0.00622 & 0.00607 \\
% $\mathrm{MAE}_{ZX}$
% & 0.00500 & 0.00499
% & 0.00628 & 0.00578 \\
% \end{tabular}
% \end{ruledtabular}
% \caption{Comparison of RIAE values, computation times, and mean absolute errors of the $XY$, $YZ$, and $ZX$ coherence estimates for SGVB and HMC applied to the \texttt{noise4a} and \texttt{noise5a} datasets.
% \avi{Simplify the table -- use the nicematrix package, we dont need all these lines}
% }
% \label{tab:lisa_results}
% \end{table*}

\begingroup
\begin{table}[t]
\centering
\renewcommand{\arraystretch}{1.18}
\setlength{\tabcolsep}{4.5pt}

\begin{NiceTabular}{l c r r r r}[
  colortbl-like,
  cell-space-limits=1pt
]
\CodeBefore
  \rowcolors{3}{gray!7}{white}
\Body

&& \Block{1-2}{\bfseries \texttt{noise4a}}
&& \Block{1-2}{\bfseries \texttt{noise5a}}
& \\[-3pt]

\cmidrule(lr){3-4}
\cmidrule(lr){5-6}

\bfseries Metric
& \bfseries Scale
& \bfseries SGVB
& \bfseries HMC
& \bfseries SGVB
& \bfseries HMC \\

\midrule

RIAE
& $10^{-2}$
& 2.667 & 2.662
& 3.37 & 3.33 \\

Time
& --
& 7.9\,min & 4.93\,h
& 5.2\,min & 2.29\,h \\

$\mathrm{MAE}_{XY}$
& $10^{-3}$
& 6.09 & 6.14
& 5.31 & 5.38 \\

$\mathrm{MAE}_{YZ}$
& $10^{-3}$
& 5.80 & 5.61
& 6.22 & 6.07 \\

$\mathrm{MAE}_{ZX}$
& $10^{-3}$
& 5.00 & 4.99
& 6.28 & 5.78 \\

\end{NiceTabular}

\caption{
Comparison of RIAE, pairwise-coherence MAE, and computation time for SGVB
and HMC applied to the \texttt{noise4a} ($M=250$) and \texttt{noise5a}
($M=400$) datasets. Error metrics exclude the $\pm1$~mHz neighborhoods of
the TDI null frequencies at 0.03, 0.06, and 0.09~Hz. Displayed error values
should be multiplied by the factor in the Scale column.
}
\label{tab:lisa_results}
\end{table}
\endgroup